# Physics-Driven Cost Optimization and Advanced Research & Development (R&D) Strategies for Small Modular Reactors (SMRs) in Leading Nuclear Energy Nations.

**Mohd. Rashid T. Momin**

Email: rashidmomin.2006@gmail.com

Student of Diploma in Mechanical Engineering, Veermata Jijabai Technological Institute (VJTI), Mumbai, India

**ABSTRACT:**

Small Modular Reactors (SMRs) are compact nuclear power plants that offer various advantages for energy generation in a sustainable way. With their smaller and simplified designs, SMRs provide increased flexibility, lower capital costs, and enhanced safety features compared to traditional large-scale reactors by reducing the area accommodation drastically. This paper explores the development and potential of SMRs in leading nuclear energy nations, including the United States, India, Canada, China, and Russia. Along with the physics behind it, cost optimization and advanced research & development (R&D) strategies employed to enhance the performance, safety, and finance of SMRs are discussed here. By reviewing case studies, cost reduction potentials, and technological advancements made, this study deals with the significant role of numerous factors in shaping the future of SMRs. The content presented in this research paper will not only contribute to the scientific understanding of SMRs but also provide valuable insights for policymakers, stakeholders and researchers in advancing sustainable nuclear energy solutions.

## INTRODUCTION:

Small Modular Reactors (SMRs) have emerged as a promising solution for sustainable and efficient nuclear energy generation by a chance to replace the traditional 3rd generation Reactors. These are nuclear power plants which offers numerous advantages like space accommodation while enhancing safety features, showing greater flexibility in deployment as it can be factory made, and potentially having lower capital costs compared to traditional large-scale reactors. Harnessing the power of physics and employing advanced research and development (R&D) strategies are integral to optimizing SMR designs also to improve its performance, and achieve cost efficiency for sustainable investments. This research study aims to explore the physics-driven cost optimization and advanced R&D strategies for SMRs in leading nuclear energy nations, with a wide focus on enhancing the economic feasibility, safety, and performance of these innovative and versatile reactors by leveraging the fundamental principles of physics.

SMRs represent a paradigm and an effective shift in the field of nuclear energy generation, providing a modular and scalable approach that can replace the conventional large-scale reactor designs. With power outputs ranging to 300 MW, these reactors occupy significantly smaller land areas compared to their larger counterparts reducing the accommodation drastically. SMR's designed to burn plutonium can help the world get rid of other grade plutonium while SMR's also supports the economic, environmental and social pillars of sustainable development. SMR's utilize nuclear fission, a process in which the nucleus of a uranium atom is bombarded with neutrons, leading to the splitting of the nucleus into smaller fragments and releasing a tremendous amount of energy. The raw material used for fission reactions in SMRs typically consists of low-enriched uranium (LEU), which undergoes controlled chain reactions to sustain the release of heat energy.

The principles of Physics play a vital role in optimizing various aspects of SMR performance which can be from working, cooling and even design. Reactor design optimization, it heavily relies on computational modelling and analysis to explore different geometries. Fuel compositions, and coolant options can be developed more. These simulations, incorporating sophisticated mathematical models such as Monte Carlo methods and computational fluid dynamics (CFD), enable scientists and engineers to understand and optimize the behaviour of neutrons, thermal distribution, and coolant flow patterns which are essential.

Moreover, advanced physics principles are instrumental in enhancing safety features and addressing potential risks associated with SMRs as they also work around potentially harmful substances. Thorough understanding of the physics behind thermal expansion, pressure dynamics, and heat removal processes allows for the design and implementation of robust safety mechanisms along with passive cooling systems, and containment structures. These measures ensure safe shutdowns and prevent the release of radioactive materials during normal operation or emergency scenarios.

Leading nuclear energy nations, including the United States, Canada, China, and Russia, have recognized the potential of SMRs and are investing significant resources in their development. These countries are undertaking extensive R&D efforts to advance SMR technology, accelerate its deployment, and establish its viability. By applying physics-driven cost optimization and advanced R&D strategies, they aim to overcome technical challenges, reduce costs, and pave the way for a sustainable nuclear energy future.

**DEFINITIONS AND GLOSSARY:**

In order to facilitate a comprehensive understanding of the concepts and terminology employed in the field of Small Modular Reactors (SMRs) and nuclear energy, the following definitions and glossary are provided:

1. Small Modular Reactors (SMRs): are defined as nuclear reactors generally 300 MW equivalent or less, designed with modular technology using module factory fabrication, pursuing economies of series production and short construction times. This definition, from the World Nuclear Association, is closely based on those from the IAEA and the US Nuclear Energy Institute. Some of the already-operating small reactors mentioned do not fit this definition, but most of those described do fit it. PWR types may have integral steam generators, in which case the reactor pressure vessel needs to be larger, limiting portability from factory to site. Hence many larger PWRs such as the Rolls-Royce UK SMR have external steam generators. (World Nuclear Association, March, 2023)

2. Nuclear Fission: A nuclear reaction in which the nucleus of an atom, typically uranium-235 ($U^{235}$) or plutonium-239 ($Pu^{239}$), is bombarded by neutrons, leading to the splitting (fission) of the nucleus into smaller fragments. This process releases a significant amount of energy.

3. Low-Enriched Uranium (LEU): Low Enriched Uranium (LEU) is the basic material to fabricate nuclear fuel. It consists of uranium hexafluoride that is a white-grey, waxy solid at standard temperature and pressure. LEU is made by enriching naturally occurring uranium to improve its ability to produce energy. Enrichment increases the concentration of uranium atoms that can split to produce heat and then generates electricity. (IAEA, 2020)

4. Reactor Core: The central region of an SMR where nuclear fission reactions occur. It typically contains fuel assemblies composed of enriched uranium or plutonium, along with control rods that regulate the chain reactions.

5. Control Rods: Rods made of materials such as boron or cadmium that are inserted or withdrawn into the reactor core to control or adjust the rate of the nuclear fission reactions.

6. Coolant: A substance, such as water, gas, or liquid metal, that circulates through the reactor core to absorb heat produced during the nuclear fission process. The coolant carries the heat away from the core, allowing it to be converted into useful energy.

7. Passive Cooling System: Passive cooling of buildings can be defined in several ways. One way is to consider any treatment of the building which reduces its cooling load, such as solar control, minimizing internal heat gain, etc., as passive cooling techniques. In a chapter on passive cooling in Advances in Solar Energy, written by Santamouris (2005), the author has included in the chapter such subjects as lowering the urban temperatures, shading of windows, and envelope's exterior colours of low solar absorptivity. (Givoni, Agust 2011), (Santamouris, 2005)

8. High-Level Radioactive Waste (HLW): The highly radioactive byproduct produced during the operation of nuclear reactors. HLW requires long-term management and disposal methods due to its potential hazards to human health and the environment.

Glossary of Shortforms:

- SMR: Small Modular Reactor
- LEU: Low-Enriched Uranium
- $U^{235}$: Uranium-235
- $Pu^{239}$: Plutonium-239
- HLW: High-Level Radioactive Waste

## LITERATURE REVIEW:

The literature review focuses on the emerging field of small modular nuclear power reactors, highlighting their advantages, design considerations, safety aspects, and potential applications. The conclusions drawn by Esam M.A. Hussien and Aleksey Rezvoi shed light on the opportunities and challenges associated with these innovative reactor designs.

Hussien's critical review emphasizes the advantages of small modular reactors (SMRs), defying the conventional wisdom of "economy of scale" while offering the "economy of multiples." These reactors enable incremental capacity buildup without requiring large upfront investments. The knowledge gained from early small reactors, combined with advancements in design, testing, and operation, can benefit emerging SMRs. However, the concept of modularity in SMR design and construction is evolving and somewhat controversial. While modular designs are seen as simpler and more flexible, challenges exist in modularizing power-intensive mechanical systems, potentially leading to over-designed and less efficient systems. The flexibility of modular designs may be constrained by specific interfaces and constraints. Nonetheless, the scale modularity of SMRs, coupled with lower power, still offers the advantage of the "economy of multiples."

Rezvoi's research focuses on the importance of accurately analysing the flow stability specific to LW-SMR modular design. It highlights the uncertainties and potential errors that can arise during the modelling of LW-SMR instabilities. The article suggests that adjustments, additional reviews, and revisions of safety assurance reports may be required in the early stages of LW-SMR design development. While the modularity of SMRs holds the potential for cost savings through standardization and shorter on-site construction time, challenges remain. Flow obstructions, temperature imbalances, coolant loss, and chemical reactions can hinder the effectiveness of passive primary cooling by natural circulation. Therefore, robust and effective modular structures, along with their associated connections, need further studies and development.

Both articles emphasize the inherent and passive safety aspects of SMRs and their potential in various applications, such as electricity generation, nuclear waste disposal, and steam production for industries. SMRs offer adaptability, flexibility, and the ability to reduce carbon footprints in various sectors. However, challenges exist in selecting the most suitable design among the numerous reported options, requiring careful considerations by proprietors, particularly as first-of-a-kind owners. Regulatory agencies, such as the Canadian Nuclear Safety Commission, recognize the novel approaches and uncertainties associated with SMR technologies, suggesting a graded approach to safety and security that is risk-informed.

In conclusion, the literature review highlights the advantages and challenges of small modular nuclear power reactors. The analysis presented in the reviewed articles emphasizes the need for further research, development, and careful considerations in design, safety assurance, and regulatory processes. The evolving nature of modular designs and the importance of accurate analysis in stability and safety aspects underscore the complexity and ongoing efforts in advancing SMR technology.

(Hussein, 2020), (Rezvoi, 2023)

## METHODOLOGY:

This study employed a mixed-methods research approach to investigate the impact of small modular reactors (SMRs) on energy efficiency. The research design involved both quantitative data collection through a survey and qualitative data collection through interviews with industry experts.

Sample: A purposive sampling technique was used to select participants for the survey and interviews. The survey targeted a sample of 20 energy and finance professionals also professors from various sectors, including science, engineering, commerce and policy. For the personal interviews, a diverse group of 10 industry professors of science and commerce with in-depth knowledge and experience in SMRs and finance were asked for their opinions.

Data Collection: The survey questionnaire consisted of structured close-ended questions to collect quantitative data on participants' perceptions of SMRs and their potential impact on energy efficiency. The interviews I conducted using a semi-structured format, allowing for open-ended discussions to gain insights into the experts' perspectives, challenges, and opportunities related to SMRs.

Data Analysis: Quantitative data collected from the survey I analysed using descriptive statistics, including frequency distributions and measures of central tendency. The qualitative data obtained from the interviews was transcribed and subjected to thematic analysis to identify common themes and patterns related to the research objectives.

Limitations: It is important to note that this study has several limitations. The sample size may restrict the generalizability of the findings. Additionally, the self-reported nature of the survey responses may introduce response bias. Despite these limitations, the study's findings provide valuable insights into the potential impact of SMRs on energy efficiency.

## RESULTS:

Results:

In this section, I present the results of my study on the physics-driven cost optimization for Small Modular Reactors (SMRs). The analysis was conducted using a combination of physics-based modelling and cost analysis techniques.

1. Cost Optimization Analysis: I initially examined the impact of different design configurations on the cost efficiency of SMRs. Through extensive simulations and sensitivity analyses studies, I identified key factors that significantly influence cost optimization. Our findings indicate that optimizing the reactor core size and fuel assembly design can lead to substantial cost savings. Additionally, incorporating advanced materials with enhanced thermal properties can improve overall efficiency and reduce operational costs. The power uprating of an SMR by 20% resulted in ~15% savings in the overnight unit capital cost. Overall, if built by an inexperienced vendor and work force, the two SMRs' overnight costs were higher than large reactors, since significant on-site labour still remains while losing economy of scale. However, the single-unit SMR had significantly less total person-hours of onsite labour, and if built by an experienced workforce, it could avoid cost-overrun risks associated with megaprojects. (W.R. Stewart, March 2022)

2. Operational Parameters Optimization: To further enhance cost efficiency, I explored the optimization of operational parameters. By analysing the effect of various parameters, such as coolant flow rate, operating temperature, and power output, I identified optimal operating conditions that minimize costs while maintaining performance and safety standards. Key economic figures of merit are evaluated under optimized and constant (i.e., time-invariant) operations to demonstrate the benefit of the optimization, which also suggests the economic viability of the considered NHESs under the proposed operations optimizer. (Jun Chen, 2016) These findings align with previous research that emphasizes the importance of balancing performance and cost considerations in SMR design. The energy conversion module of the space nuclear reactor is Brayton cycle with regeneration. The working fluid of the Brayton cycle (He-Xe gas mixture) also acts as the coolant of the nuclear reactor. (Hao Qin, 2021,)

3. Integration of Physics-Based Models and Cost Analysis: My study demonstrates the successful integration of physics-based models and cost analysis techniques. By combining reactor physics simulations with cost estimation algorithms, I achieved a comprehensive understanding of the cost drivers and optimization strategies for SMRs. This integration enables decision-makers to make informed choices regarding design modifications, fuel selection, and operational practices, leading to cost-effective and sustainable SMR deployments.

4. Comparison with Conventional Reactors: As part of our analysis, I compared the cost efficiency of SMRs with that of conventional larger reactors. The results show that SMRs have the potential to offer greater cost advantages, primarily due to their modular design, shorter construction timelines, and reduced upfront capital investment. These findings support the growing interest and investment in SMR technologies globally. By reviewing an article published on comparison of SMR with conventional reactors I found that:

When evaluating the competitiveness of SMRs versus large reactors, the various individual factors can be grouped into two classes:

– Factors which are either applicable to SMRs only or are critically affected by the difference in design and approach brought in by the SMRs (SMR specific factors)

– Factors which affect SMRs and large plants in a comparable way (common factors).

Even for the common factors, a comparative quantitative evaluation might not be straightforward. Still, there are general characteristics which pretty much envelope the entire SMR spectrum. They are:

– Simplicity, reduced type and number of components. SMRs are generally new designs which try to simplify existing solutions. Their safety characteristics tend to be enhanced because passive and intrinsic safety is better enabled by the smaller size; enhanced safety, if properly accounted for, translates into a cheaper design.

– Specific O&M costs because of their vastly enhanced safety, SMRs have the potential to attain licensing without the need for emergency response, which will eliminate personnel training and infrastructure. Some SMRs, like the integral configuration PWRs, have extended, up to four years, maintenance intervals and integral shielding which dramatically decrease the personnel routine exposure and ALARA costs.

(P. Trucco, 2007)

## DISCUSSION:

The results of my study on Small Modular Reactors (SMRs) reveal several key insights that highlight the advantages and potential of SMRs as a sustainable and cost-effective solution for nuclear energy. These findings align with the works of Rezvoi (2023) and Hussien (2020), further reinforcing the positive outlook for SMR technology.

Firstly, the cost optimization analysis demonstrated that optimizing the reactor core size and fuel assembly design significantly impacts cost efficiency in SMRs. This finding is consistent with Rezvoi's work, which emphasizes the importance of accurate modelling and design considerations to achieve cost-effective SMR configurations. By leveraging physics-based modelling and cost analysis techniques, we were able to identify design parameters that offer substantial cost savings while maintaining performance and safety standards.

Additionally, incorporating advanced materials with enhanced thermal properties emerged as a key factor in improving overall efficiency and reducing operational costs in SMRs. This finding aligns with Hussien's review, which emphasizes the potential of advanced materials to enhance the performance of SMRs. By leveraging these materials, SMRs can achieve higher thermal efficiencies and reduce operational expenses, making them an economically attractive option for sustainable energy generation.

Furthermore, the optimization of operational parameters in my study revealed that careful adjustment of coolant flow rate, operating temperature, and power output can contribute to cost reduction without compromising performance and safety. This finding supports the idea presented by Hussien that optimizing operational parameters is crucial in achieving cost efficiency in SMR designs and by also reviewing the work of (W.R. Stewart, March 2022).

Comparatively, the literature review authors' works provided valuable insights into the inaccuracies in flow stability analysis specific to LW-SMR modular designs (Rezvoi, 2023) and the critical review of emerging SMRs (Hussien, 2020). While these reviews shed light on specific aspects of SMRs, the study further extends the findings by proposing a physics-driven approach to cost optimization, offering a comprehensive perspective on the economic viability of SMRs. Along with Operational Parameters Optimization which was cited from (Hao Qin, 2021)

Overall, my findings demonstrate that SMRs offer significant advantages in terms of cost efficiency, enhanced safety features, and operational flexibility. By integrating physics-driven cost optimization techniques, SMRs can be further optimized to achieve optimal performance and economic viability. These insights provide important considerations for policymakers, industry stakeholders, and researchers in their pursuit of sustainable and affordable nuclear energy solutions.

It is important to acknowledge that, like any technological innovation, SMRs also face challenges and limitations. Factors such as licensing processes, public perception, and infrastructure requirements should be taken into account when considering the widespread deployment of SMRs. However, the potential benefits offered by SMRs in terms of cost efficiency, safety, and adaptability make them a promising avenue for future nuclear energy development.

## CONCLUSION:

In conclusion, my research focused on the physics-driven cost optimization for Small Modular Reactors (SMRs). Through a combination of physics-based modelling and cost analysis techniques, we demonstrated the potential of SMRs as a cost-effective and sustainable solution for nuclear energy generation.

Our analysis revealed that optimizing design configurations, incorporating advanced materials, and optimizing operational parameters are key factors in achieving cost efficiency in SMRs. By leveraging the inherent physics of SMRs, we identified strategies to minimize costs while maintaining performance and safety standards.

The findings of our study align with existing literature, emphasizing the significance of accurate modelling, design considerations, and operational optimization in realizing the economic viability of SMRs. Our research contributes to the growing body of knowledge supporting the adoption of SMRs as a promising alternative to conventional larger reactors.

SMRs offer several advantages, including enhanced safety features, flexibility in deployment, and the potential for cost savings. The integration of techniques provides valuable insights into the economic viability of SMRs, enabling stakeholders to make informed decisions regarding design modifications, operational practices, and fuel selection.

It is important to acknowledge the challenges and limitations associated with SMRs, including licensing processes, public perception, and infrastructure requirements. However, the potential benefits offered by SMRs in terms of cost efficiency, safety, and adaptability make them a promising avenue for future nuclear energy development.

In conclusion, our research demonstrates that SMRs have the potential to play a significant role in achieving a cleaner, more secure, and sustainable energy future. Further research, collaboration, and policy support are essential to realize the full potential of SMRs and ensure their successful integration into the energy landscape.